\documentclass[amsmath,amssymb,floatfix,aps,twocolumn,superscriptaddress,nofootinbib,notitlepage,prl]{revtex4-2}

\usepackage{amsmath,amsthm,amssymb,bm,graphicx,float,xcolor,changes,mathtools,enumitem}
\usepackage[colorlinks=true,urlcolor=purple,linkcolor=blue,citecolor=green,bookmarks=false]{hyperref}
\allowdisplaybreaks

\newcommand{\be}{\begin{equation}}
\newcommand{\ee}{\end{equation}}
\newcommand{\6}{\partial}

\usepackage{pdfpages} 
\makeatletter
\AtBeginDocument{\let\LS@rot\@undefined}
\makeatother

\begin{document}

\title{Dynamical fermionization  in one-dimensional spinor gases at finite temperature}

\author{Ovidiu I. P\^{a}\c{t}u}
\affiliation{Institute for Space Sciences, Bucharest-M\u{a}gurele, R 077125, Romania}

\begin{abstract}

Following the removal of  axial confinement the momentum distribution of a Tonks-Girardeau gas approaches that of a system of
noninteracting spinless fermions in the initial harmonic trap. This phenomenon, called dynamical fermionization, has been
experimentally confirmed in the case of the Lieb-Liniger model and theoretically predicted in the case of multicomponent
systems at zero temperature. We prove analytically that for all  spinor gases with strong repulsive contact interactions at
finite temperature the momentum distribution after release from the  trap asymptotically approaches that of a system of
spinless fermions at the same temperature but with a renormalized chemical potential  which depends on the number of
components of the spinor system.   In the case of the Gaudin-Yang model we check numerically our analytical predictions using
the results obtained  from a  nonequilibrium generalization of  Lenard's formula describing the time evolution of the
field-field correlators.
\end{abstract}

\maketitle

{\textit{Introduction.---}}
In the last decade considerable effort has been devoted to understanding the nonequilibrium dynamics of one-dimensional (1D)
integrable and near-integrable many-body systems after the realization that such systems  do not thermalize \cite{KWW07,RDO08,
PSSV11,KLG21}. This flurry of activity resulted in the introduction of  powerful techniques like the quench action \cite{CE,
Caux1}, generalized hydrodynamics \cite{CDY,BCNF} and in the investigation of various nonequilibrium scenarios in both single
component \cite{PB07,JPGB08,CEF11,IA12,NWBC14, Poz13,WNBF14,PMWK14,INWCE,PPV16,BWE16,CSC13a,RS14,KCTC17,CKC18,VIR17,AGBKa,ABGKb,ASK19,P20,
SBDD19,RBD19, RCDD20,MZD21,SKCD21} and multi-component systems \cite{YP16,IN17,MBPC17,ZVR19,SMS20,WYC20,NT20,NT21,SCP21,RBC22,SCP22}.

At zero temperature the momentum distribution of 1D strongly interacting bosons released from a harmonic trap will asymptotically
approach the momentum distribution of a similar number of spinless fermions in the initial trap. This phenomenon, dubbed dynamical
fermionization (DF) was theoretically  predicted in \cite{RM05,MG05} (see also \cite{delC,GP08,BHLM12,CGK15,XR17,CDD17}) and
experimentally confirmed recently using  ultracold atomic gases \cite{WML20}. DF was also theoretically predicted to occur in
multicomponent systems, bosonic, fermionic \cite{ASYP21} or mixtures \cite{P22}  using the  factorization of the wavefunctions in
charge and spin components in the strongly interacting regime \cite{OS90,IP98,DFBB08,GCWM09,VFJV14,LMBP15,YC16,DBBR17,YAP22}. At
finite temperature results in the literature regarding DF are almost nonexistent with the only example that we are aware of being
the numerical confirmation in the case of single component bosons reported in \cite{XR17}. Generalizing the method of \cite{MG05}
for finite temperature it can be shown \cite{SM}  that for a system of  trapped impenetrable bosons  described by the grandcanonical
ensemble at temperature $T$ and chemical potential $\mu$ that DF is present and the asymptotical momentum distribution is the same
as the one for a system of spinless fermions at the same temperature and chemical potential. The situation in the case of
multicomponent systems is, obviously, more complicated. Naively, one would expect that if DF occurs in a multicomponent system at
finite temperature  then the asymptotic  momentum distribution would be expressed as a sum of momentum distributions of free
fermions with different chemical potentials. Contrary to this expectation in this paper we show that for a spinor system  at finite
temperature the asymptotic momentum distribution after release from the trap approaches that of a system of spinless fermions at the
same temperature but with a  \textit{renormalized chemical potential}, denoted by $\mu'$, which depends on the number of components
of the system (or magnetic field in the case of unbalanced systems) but not on the statistics of the particles.
More precisely, for any harmonically trapped  multicomponent gas, bosonic or fermionic, with strong  repulsive contact interactions
we will show that after release from the trap:
(0) the initial density  profile of the spinor gas is the same as  the density profile of spinless noninteracting fermions described
by $T$ and $\mu'$ (this is in general called fermionization);
(1) the  asymptotic momentum distribution has the same shape as the initial density profile; and
(2) the  asymptotic momentum distribution is the same as the one for spinless noninteracting fermions characterized by $T$ and $\mu'$
which represents the dynamical fermionization of the gas.
In the case of the Gaudin-Yang model we present results also for each component (spin-up and spin-down) and numerically check our
analytical predictions by deriving  an extremely efficient determinant representation for the correlators which can be understood as
the nonequilibrium multicomponent generalization of Lenard's formula \cite{L66}.

\textit{The Gaudin-Yang model.---}
It is instructive to look first at the two-component case which provides the general template for the proof of DF in spinor gases but
also has the advantage of allowing one to investigate the contribution of each component (and not  only the sum like in the general case)
both analytically and numerically. The Gaudin-Yang model \cite{Ga1,Y1}  describes one-dimensional fermions or bosons with contact
interactions and is the natural two-component generalization of the  Lieb-Liniger model \cite{LL63}. In the presence of a
time-dependent  harmonic potential  $V(x,t)=m\omega^2(t) x^2/2$  the Hamiltonian reads
\begin{align}\label{Hc}
\mathcal{H}=&\int dx\,  \frac{\hbar^2}{2m} (\6_x \Psi^\dagger \6_x \Psi)+g \, :(\Psi^\dagger \Psi)^2:\nonumber\\
&\qquad +(V(x,t)-\mu)(\Psi^\dagger \Psi)+B(\Psi^\dagger\sigma_z\Psi)\, ,
\end{align}
where $\Psi=\left(\begin{array}{c} \Psi_\uparrow(x)\\ \Psi_\downarrow(x)\end{array}\right)\, ,$ $ \Psi^\dagger=\left(\Psi_\uparrow^
\dagger(x),\Psi_\downarrow^\dagger(x)\right)\, ,$ $\sigma_z$ is the third Pauli matrix, $\mu$ is the chemical potential, $B$ the magnetic
field and $:\  :$ denotes normal ordering.  $\Psi_{\uparrow,\downarrow}(x)$ are fermionic or bosonic fields which satisfy the commutation
relations $\Psi_\alpha(x)\Psi_ \beta^\dagger(y)-\varepsilon\Psi_\beta^\dagger(y)\Psi_\alpha(x)=\delta_{\alpha\beta}\delta(x-y)$ with
$\varepsilon=1$ in the bosonic case  and $\varepsilon=-1$ in the fermionic case. In this paper we will investigate the nonequilibrium dynamics in
the Tonks-Girardeau (TG) regime characterized by $g=\infty$. In the TG regime, also known as the impenetrable regime,
the system is integrable even in the presence of  the external potential and at $t=0$ the eigenstates
of a system  of $N$ particles of which $M$ have spin-down are [$\boldsymbol{x}=(x_1,\cdots,x_N)$, $d\boldsymbol{x}=\prod_{i=1}^N dx_i$]
\begin{align}\label{eigen}
|\Phi_{N,M}(\boldsymbol{j},\boldsymbol{\lambda})\rangle=&\int d\boldsymbol{x}\sum_{\alpha_1,\cdots,\alpha_N=
\{\downarrow,\uparrow\}} \chi_{N,M}^{\alpha_1\cdots \alpha_N}(\boldsymbol{x}|\boldsymbol{j},\boldsymbol{\lambda})\nonumber\\
&\qquad\qquad\Psi_{\alpha_N}^\dagger(x_N)\cdots\Psi_{\alpha_1}^\dagger(x_1)|0\rangle\, .
\end{align}
Here the summation is over the $C^N_M$ sets of $\alpha$'s of which $M$ are spin-down and $N-M$ are spin-up and $|0\rangle$ is the Fock
vacuum satisfying  $\Psi_\alpha(x)|0\rangle=\langle0| \Psi_\alpha^\dagger(x)=0$ for all $x$ and $\alpha$. The eigenstates (\ref{eigen}) are
identified by two sets of unequal numbers $\boldsymbol{j}=(j_1,\cdots,j_N)$ and  $\boldsymbol{\lambda}=(\lambda_1,\cdots, \lambda_M)$ which
correspond to the charge  and  spin degrees of freedom. The normalized wavefunctions are
\begin{align}\label{wfall}
\chi_{N,M}^{\alpha_1\cdots\alpha_N}(\boldsymbol{x}|\boldsymbol{j},\boldsymbol{\lambda})=&\frac{1}{N!\,N^{M/2}}
\left[\sum_{P\in S_N}(-\varepsilon)^P\eta_{N,M}^{\alpha_{P_1}\cdots\alpha_{P_N}}(\boldsymbol{\lambda})\right.\nonumber\\
&\qquad\ \ \ \ \  \times\theta(P\boldsymbol{x}) \bigg ]
\det_N\left[\phi_{j_a}(x_b)\right]\, ,
\end{align}
with the determinant expressed in terms of Hermite functions of frequency $\omega_0=\omega(t\le 0)$ i.e., $ \phi_j(x)= \left(2^j j !\right
)^{-1/2} \left(\frac{ m\omega_0}{\pi \hbar}\right)^{1/4} e^{-\frac{m\omega_0 x^2}{2\hbar}} H_j\left(\sqrt{\frac{m\omega_0}{\hbar}}x\right)$
with $H_j(x)$ the Hermite polynomials. In (\ref{wfall}) the sum is over the permutations of $N$ elements and $\theta(P\boldsymbol{x})=\theta
(x_{P_1}<\cdots< x_{P_N})= \prod_{j=2}^N\theta(x_{P_j} -x_{P_{j-1}})$ with $\theta(x)$ the Heaviside function. The $\eta_{N,M}$ functions
describing the spin sector are the wavefunctions of the XX spin-chain with periodic boundary conditions
$
\eta_{N,M}^{\alpha_1\cdots\alpha_N}(\boldsymbol{\lambda})=\prod_{j>k}\mbox{sign}(n_j-n_k)\det_M\left(e^{in_a\lambda_b}\right)\, ,
$
where $\boldsymbol{\lambda}=(\lambda_1,\cdots,\lambda_M)$ with $e^{i\lambda_a N}=(-1)^{M+1}$  and $\boldsymbol{n}=(n_1,\cdots, n_M)$ is a set
of integers, $n_a\in\{1,\cdots,N\}$, describing the positions of the spin-down particles in the ordered set $\{x_1,\cdots, x_N\}$.
The wavefunctions (\ref{wfall}) represent the natural generalization of the Bethe ansatz wavefunctions for the Gaudin-Yang model
\cite{IP98} in the presence of an external confining potential. They solve the many-body Schr\"odinger equation,
have the appropriate symmetries when exchanging two particles
of the same type, satisfy the hard-core condition (the wavefunctions vanish when two coordinates are equal) and form a complete system.
We stress again that the wavefunctions (\ref{wfall}) and all the results derived below are valid only in the TG regime ($g=\infty$).
The eigenstates (\ref{eigen}) are normalized
$
\langle\Phi_{N',M'}(\boldsymbol{j'},\boldsymbol{\lambda'}) |\Phi_{N,M}(\boldsymbol{j},\boldsymbol{\lambda})\rangle
=\delta_{N'N}\delta_{M'M}\delta_{\boldsymbol{j'} \boldsymbol{j'}}\delta_{\boldsymbol{\lambda'} \boldsymbol{\lambda}}\, ,
$
highly degenerate (their energies do not depend on $\boldsymbol{\lambda}$) and satisfy $\mathcal{H}|\Phi_{N,M}(\boldsymbol{j}, \boldsymbol{
\lambda})\rangle=E_{N,M}(\boldsymbol{j},\boldsymbol{\lambda})|\Phi_{N,M} (\boldsymbol{j},\boldsymbol{\lambda})\rangle$ with $E_{N,M}(
\boldsymbol{j},\boldsymbol{\lambda})=\sum_{i=1}^N\left[\hbar \omega_0(j_i+1/2)-\mu+B\right]-2BM$. It should be noted that the energy spectrum
is independent of statistics.

\textit{Quench protocol.---}
We are interested in investigating the dynamics of the real space and momentum densities at finite temperature after release from the
trap. Our quench protocol is the following. Initially the system  is prepared in a grandcanonical thermal state with the density matrix
\begin{align}\label{e10}
\boldsymbol{\rho}^{\mu,B,T}=&\sum_{N=0}^\infty \sum_{M=0}^N\sum_{\{\boldsymbol{j}\}} \sum_{\{\boldsymbol{\lambda}\}}
p_{N,M}^{\boldsymbol{j},\boldsymbol{\lambda}}(\mu,B,T)\nonumber\\
&\qquad\qquad\times|\Phi_{N,M}(\boldsymbol{j},\boldsymbol{\lambda})\rangle\langle\Phi_{N,M}(\boldsymbol{j},\boldsymbol{\lambda})|\, ,\ \
\end{align}
where $p_{N,M}^{\boldsymbol{j},\boldsymbol{\lambda}}(\mu,B,T)=e^{-E_{N,M}(\boldsymbol{j},\boldsymbol{\lambda})/k_B T}/\mathcal{Z}(\mu,B,T)$,
$\mathcal{Z}(\mu,B,T)=\mbox{Tr}[e^{-\mathcal{H}^I/k_B T}]$ is the partition function of the Gaudin-Yang model and $\mathcal{H}^I$  is
the Hamiltonian (\ref{Hc}) at $t=0$ ($\omega(t\le 0)=\omega_0$). At $t>0$ we remove the axial confinement  and the system evolves with
$\mathcal{H}^F$ which is the Hamiltonian (\ref{Hc}) with $\omega(t>0)=0$. Our main objects of study are the field-field correlators  defined as ($\sigma=\{\uparrow,
\downarrow\}$)
\be\label{e11}
g^{\mu,B,T}_\sigma(\xi_1,\xi_2;t)=\mbox{Tr}\left[\boldsymbol{\rho}^{\mu,B,T} \Psi^\dagger_\sigma(\xi_1,t)\Psi_\sigma(\xi_2,t)\right]\, ,
\ee
with $\Psi^\dagger_\sigma(\xi,t)=e^{i \mathcal{H}^F t}\Psi^\dagger_\sigma(\xi)e^{-i \mathcal{H}^F t}$. From the correlators one can obtain
the real space densities $\rho^{\mu,B,T}_\sigma(\xi,t)= g^{\mu,B,T}_\sigma(\xi,\xi;t)$ and the momentum distributions $n^{\mu,B,T}_\sigma(p,t)
=\int e^{i p(\xi_1-\xi_2)/\hbar}g^{\mu,B,T}_\sigma(\xi_1,\xi_2;t) \, d\xi_1 d\xi_2/2\pi .$ Because  $g^{\mu,B,T}_\uparrow(\xi_1,\xi_2;t)=
g^{\mu,-B,T}_\downarrow(\xi_1,\xi_2;t)$ it is sufficient to consider only one of the correlators.

\textit{Time-evolution of the correlators.---}
The important observation which allows for the analytical investigation  of the dynamics is that the spin component of the wavefunctions
remains frozen during the time-evolution due to the strong interactions between the particles \cite{ASYP21,P22}. The charge component of
the wavefunctions (\ref{wfall}) is expressed in terms of harmonic oscillator functions whose dynamics in the case of time dependent frequency
is known \cite{PP70,PZ}  and is implemented by the scaling transformation $\phi_j(x,t)=\frac{1}{\sqrt{b}}\phi_j\left(\frac{x}{b},0\right)
\exp\left[i\frac{m x^2}{2\hbar} \frac{\dot{b}} {b}-i E_j\tau(t)\right]$ with $E_j=\hbar\omega_0(j+1/2)$ and  $\tau(t)=\int_0^tdt'/b^2(t')$.
In the previous relations $b(t)$ is a solution of the Ermakov-Pinney equation $\ddot{b}=-\omega(t)^2b+\omega_0^2/b^3$ with boundary
conditions $b(0)=1$, $\dot{b}(0)=0$. Therefore, we can investigate the dynamics computing the correlators at $t=0$ and then applying the
scaling transformation. At $t=0$ the correlators in the initial thermal state described by the density matrix (\ref{e10}) can be written as
\be\label{e12}
g^{\mu,B,T}_\sigma(\xi_1,\xi_2)=\sum_{N=1}^\infty\sum_{M=0}^N\sum_{\{\boldsymbol{j}\}}\sum_{\{\boldsymbol{\lambda}\}}
p_{N,M}^{\boldsymbol{j},\boldsymbol{\lambda}}\,
G_{N,M,\sigma}^{\boldsymbol{j},\boldsymbol{\lambda}}(\xi_1,\xi_2)\, ,
\ee
with $G_{N,M,\sigma}^{\boldsymbol{j},\boldsymbol{\lambda}}(\xi_1,\xi_2)=\langle\Phi_{N,M}(\boldsymbol{j},\boldsymbol{\lambda})|
\Psi^\dagger_\sigma(\xi_1)\Psi_\sigma(\xi_2)|\Phi_{N,M}(\boldsymbol{j},\boldsymbol{\lambda})\rangle$. The $G$  functions are the normalized
mean values of bilocal operators in arbitrary states described by $\boldsymbol{j}$ and $\boldsymbol{\lambda}$.
Introducing a new parametrization  \cite{ID06,P22,SM} which makes the  decoupling of the degree of freedom
explicit then, for $\xi_1\le \xi_2$, the $G$ functions can be expressed as sums of products of spin and charge functions (for their explicit
expressions see \cite{SM})
\begin{align}\label{e13}
G_{N,M,\sigma}^{\boldsymbol{j},\boldsymbol{\lambda}}(\xi_1,\xi_2)&=\frac{1}{c_\sigma N^M}\sum_{d_1=1}^N\sum_{d_2=d_1}^N S_\sigma(d_1,d_2)
\nonumber\\ &\qquad\qquad\qquad\times I(d_1,d_2;\xi_1,\xi_2)\, ,
\end{align}
with $c_\downarrow=(N-M)!  (M-1)!$ and $c_\uparrow=(N-M-1)! M!$.
The time-evolution of the correlators  is obtained by plugging the scaling transformation of the Hermite functions
in the  expression for $G_{N,M,\sigma}^{\boldsymbol{j},\boldsymbol{\lambda}}(\xi_1,\xi_2)$ in terms of wavefunctions (see \cite{SM}).
We find  ($l_{0}=\sqrt{\hbar/(m\omega_0)}$)
\be\label{e5}
G_{N,M,\sigma}^{\boldsymbol{j},\boldsymbol{\lambda}}(\xi_1,\xi_2;t)=\frac{1}{b}G_{N,M,\sigma}^{\boldsymbol{j},\boldsymbol{\lambda}}
\left(\frac{\xi_1}{b},\frac{\xi_2}{b};0\right)e^{-\frac{i}{b}\frac{\dot b}{\omega_0}
\frac{\xi_1^2-\xi_2^2}{2l_{0}^2}}\, ,
\ee
and introducing the notation $\widetilde{G}_{N,M,\sigma}^{\boldsymbol{j},\boldsymbol{\lambda}}(p,t)=\int e^{i p(\xi_1-\xi_2)/\hbar}
G_{N,M,\sigma}^{\boldsymbol{j},\boldsymbol{\lambda}}(\xi_1,\xi_2;t) \, d\xi_1 d\xi_2/2\pi $ we have
\begin{align}\label{momd}
\widetilde{G}_{N,M,\sigma}^{\boldsymbol{j},\boldsymbol{\lambda}}(p,t)=&\frac{b}{2\pi}\int G_{N,M,\sigma}^{\boldsymbol{j},
\boldsymbol{\lambda}}(\xi_1,\xi_2;0) \nonumber\\
&\ \ \ \times e^{-i b\left[\frac{\dot{b}}{\omega_0}\frac{\xi_1^2-\xi_2^2}{2l_{0}^2}-
\frac{p(\xi_1-\xi_2)}{\hbar}\right]}\, d\xi_1 d\xi_2\, .
\end{align}
The dynamics of the real space density and  momentum distribution is derived using (\ref{e5}) and (\ref{momd}) in Eq.~(\ref{e12}).

\textit{Analytical derivation of dynamical fermionization.---}
As a preliminary step we will compute the partition function of the Gaudin-Yang (GY) model which appears in the definition of the state probabilities
$p_{N,M}^{\boldsymbol{j},\boldsymbol{\lambda}}(\mu,B,T)$ describing the density matrix (\ref{e10}). We should point out that the
thermodynamics of trapped impenetrable particles with contact interactions is independent of statistics (the energy spectrum is identical
and double occupancies are excluded). In the case of homogeneous systems a proof can be found in \cite{IP98}.
Using the identity $\sum_{M=0}^N
\sum_{\{\boldsymbol{\lambda}\}}e^{\frac{2BM}{k_B T}}=(1+e^{\frac{2B}{k_B T}})^N$ we obtain
$ \mathcal{Z}(\mu,B,T)=
\sum_{N=0}^\infty \sum_{\{\boldsymbol{j}\}} \left[2\cosh(B/k_B T)\right]^N e^{-E_{N}(\boldsymbol{j})/k_B T}
$
with $E_{N}(\boldsymbol{j})=\sum_{i=1}^N\left[\hbar \omega_0(j_i+1/2)-\mu\right]$ which shows that the partition function of the harmonically
trapped GY model in the TG regime is the same as the one of trapped spinless free fermions $\mathcal{Z}_{\textsf{FF}}(\mu',T)$ at the same
temperature but with renormalized chemical potential (this is the generalization of the homogeneous result first obtained by Takahashi in
\cite{T71})
\be\label{mu}
\mu'=\mu+k_BT\ln[2\cosh(B/k_B T)]\, .
\ee

Let us investigate the densities at $t=0$. From the definition (\ref{e12}) we have $\rho^{\mu,B,T}_\sigma(\xi)=\sum_{N=1}^\infty\sum_{M=0}^N
\sum_{\{\boldsymbol{j}\}}\sum_{\{\boldsymbol{\lambda}\}} p_{N,M}^{\boldsymbol{j},\boldsymbol{\lambda}}\,G_{N,M,\sigma}^{\boldsymbol{j},
\boldsymbol{\lambda}}(\xi,\xi)\, $ with $G_{N,M,\sigma}^{\boldsymbol{j},\boldsymbol{\lambda}}(\xi,\xi)=\sum_{d=1}^N S_\sigma(d,d) I(d,d;\xi,\xi)
/(c_\sigma N^M)\, .$ It can be shown \cite{SM} that $S_\downarrow(d,d)=(N-M)!M!N^{M-1}$,  $S_\uparrow(d,d)=(N-M-1)! M! N^{M-1}(N-M)$  and
that $\sum_{d=1}^N I(d,d;\xi,\xi)=G_{N,\textsf{FF}}^{\boldsymbol{j}}(\xi,\xi)$ where
$G_{N,\textsf{FF}}^{\boldsymbol{j}}(\xi,\xi)$  is the density of free fermions (in the state $\boldsymbol{j}$) at position $\xi$.
Using these results we obtain
\begin{align}\label{e15}
\rho_\downarrow^{\mu,B,T}(\xi)=\frac{e^{B/k_B T}}{2\cosh(B/k_BT)}\rho_{\textsf{FF}}^{\mu',T}(\xi)\, ,\ \
\end{align}
$\rho_\uparrow^{\mu,B,T}(\xi)=\rho_\downarrow^{\mu,-B,T}(\xi)$ and $\rho_\downarrow^{\mu,B,T}(\xi)+\rho_\uparrow^{\mu,B,T}(\xi)=
\rho_{\textsf{FF}}^{\mu',T}(\xi)$ proving that the initial densities are proportional to the densities of trapped spinless free fermions at
the same temperature and chemical potential given by (\ref{mu}) (property 0 from the introduction).

Now we can investigate the dynamics. In the case of free expansion the solution of the Ermakov-Pinney equation is $b(t)=(1+\omega_0^2 t^2)^{1/2}$
and in the large time limit we have $\lim_{t\rightarrow \infty}b(t)=\omega_0 t$ and $\lim_{t\rightarrow \infty}\dot{b}(t)=\omega_0.$
The momentum distribution is
\be
n_\sigma(p,t)=\sum_{N=1}^\infty\sum_{M=0}^N\sum_{\{\boldsymbol{j}\}}\sum_{\{\boldsymbol{\lambda}\}}p_{N,M}^{\boldsymbol{j},\boldsymbol{\lambda}}\,
\widetilde{G}_{N,M,\sigma}^{\boldsymbol{j},\boldsymbol{\lambda}}(p,t)\, ,
\ee
and we need $\lim_{t\rightarrow \infty} \widetilde{G}_{N,M,\sigma}^{\boldsymbol{j},\boldsymbol{\lambda}}(p,t)$. Using the method of stationary
 phase (Chap. 6 of \cite{BHbook} or Chap. 2.9 of \cite{Ebook})  in (\ref{momd}) with the points of stationary phase being $\xi_0=p\omega_0l_{0}^2
/(\dot{b} \hbar)$ for both integrals we find
$
\widetilde{G}_{N,M,\sigma}^{\boldsymbol{j},\boldsymbol{\lambda}}(p,t)\underset{t \rightarrow \infty}{\sim}\left|\frac{\omega_0 l_{0}^2}{\dot{b}}\right|
G_{N,M,\sigma}^{\boldsymbol{j},\boldsymbol{\lambda}} \left(\frac{p\omega_0l_{0}^2}{\dot{b} \hbar},\frac{p\omega_0l_{0}^2}{\dot{b} \hbar};0 \right)
\, .
$
We have $G_{N,M,\downarrow}^{\boldsymbol{j},\boldsymbol{\lambda}}(\xi,\xi)=M G_{N,\textsf{FF}}^{\boldsymbol{j}}(\xi,\xi)/N$ and $G_{N,M,\uparrow}^{
\boldsymbol{j},\boldsymbol{\lambda}}(\xi,\xi)=(N-M) G_{N,\textsf{FF}}^{\boldsymbol{j}}(\xi,\xi)/N.$ Performing similar computations like in the case
of the initial densities we obtain
\begin{align}\label{e16}
n_\downarrow^{\mu,B,T}(p,t)\underset{t \rightarrow \infty} {\sim}  l_0^2\, \frac{e^{B/k_B T}}{2\cosh(B/k_BT)}\,\rho_{\textsf{FF}}^{\mu',T}
\left(\frac{p l_0^2}{\hbar}\right)\, ,
\end{align}
and $n_\uparrow^{\mu,B,T}(p,t)=n_\downarrow^{\mu,-B,T}(p,t)$ which shows that the asymptotic momentum distributions have the same shape as the initial
densities (property 1). Finally, using the identity $n_{\textsf{FF}}^{\mu,T}(p)=l_0^2\, \rho_{\textsf{FF}}^{\mu,T}\left(p l_0^2/\hbar\right)$ (see
Appendix E of \cite{P22}) we obtain
\begin{align}\label{e17}
n_\downarrow^{\mu,B,T}(p,t)\underset{t \rightarrow \infty} {\sim}  \frac{e^{B/k_B T}}{2\cosh(B/k_BT)}\, n_{\textsf{FF}}^{\mu',T}\left(p\right)\, ,\ \
\end{align}
and $n_\downarrow^{\mu,B,T}(p,t)+n_\uparrow^{\mu,B,T}(p,t)\underset{t \rightarrow \infty} {\sim}n_{\textsf{FF}}^{\mu',T}\left(p\right)$ which
proves the dynamical fermionization at finite temperature (property 2).

\begin{figure}
\includegraphics[width=1\linewidth]{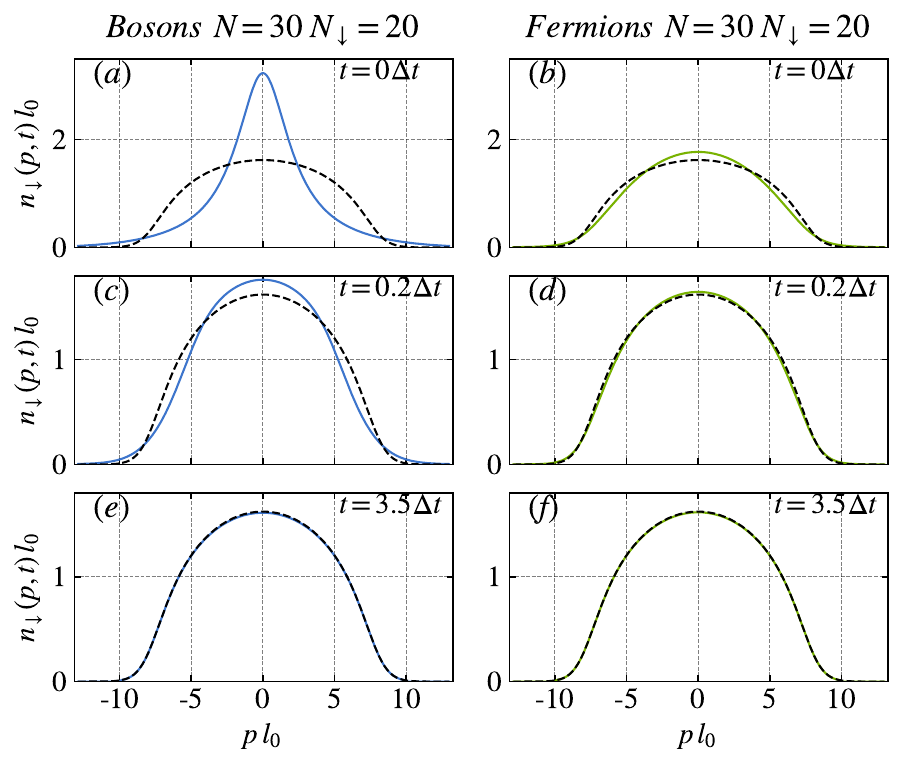}
\caption{Dynamics of the momentum distribution of spin-down particles after release from the trap in the GY model with $N=30$  and $N_\downarrow=20$.
The temperature and initial trap frequency are $T=5$ and $\omega_0=1$  ($\mu=26.22$, $|B|=1.73$, $\Delta t= \pi/\omega_0$). The continuous line in panels a), c) and e) [b), d)
and f)] represents the momentum distribution $n_\downarrow(p,t)$ for a bosonic (fermionic) system while the dashed line is the analytical prediction
Eq.~(\ref{e17}).
}
\label{fig1}
\end{figure}

We can numerically check the analytical predictions given by Eq.~(\ref{e17}) using a determinant representation for the field correlators which
represents the other main result of this paper. This representation obtained via summation of the form factors is the nonequilibrium multicomponent
generalization of Lenards's formula \cite{L66} originally introduced for impenetrable bosons  and reads [$g^{\mu,B,T}_\downarrow(\xi_1,\xi_2;t)=
g^{\mu,-B,T}_\uparrow( \xi_1,\xi_2;t)$]
\be\label{gmsmall}
g^{\mu,B,T}_\downarrow(\xi_1,\xi_2;t)=\det\left(1+\gamma \textsf{V}+\textsf{R}\right)-\det\left(1+\gamma \textsf{V}\right)\, ,
\ee
with $\gamma=-(1+e^{2 B/T}+\varepsilon)\mbox{sign}(\xi_2-\xi_1)$ and the elements of the (infinite) matrices $\textsf{V}, \textsf{R}$ are given by
$\textsf{V}_{a,b}=\sqrt{f(a)f(b)} \int_{\xi_1}^{\xi_2}\overline{\phi}_a(v,t)\phi_b(v,t)\, dv$ and $\textsf{R}_{a,b}=\sqrt{f(a)f(b)}\, \overline{
\phi}_a(\xi_1,t) \phi_b(\xi_2,t)$ where $f(a)=e^{-B/T}/[2\cosh(B/T)+e^{\frac{\hbar\omega_0(a+1/2)-\mu}{T}}]$ is the Fermi function and $\phi_a(v,t)$
are the time-evolved harmonic orbitals.
In addition to representing  the starting point for the rigorous derivation of various analytical properties of the correlators (for example one
can show that $g^{\mu,B,T}_{\downarrow,\uparrow}(\xi_1,\xi_2;t)$ can be expressed in terms of Painlev\'e transcendents) Eq.~(\ref{gmsmall}) is also
extremely efficient numerically due to the fact that the main computational effort is reduced to the calculation of partial overlaps of the single
particle evolved wavefunctions and, therefore, can be used to investigate different experimentally relevant quench scenarios like breathing
oscillations \cite{AGBKa,ABGKb}, quantum Newton's  cradle \cite{KWW07,BWE16}, periodic modulation of the frequency \cite{ASK19}, etc.,  which were
not previously accessible in the case of  multicomponent systems. Fig.~\ref{fig1} presents the dynamics of $n_\downarrow(p,t)$ derived from
(\ref{gmsmall}) for an unbalanced system with $N=30$   particles and $N_\downarrow=20$   after release from the trap  which shows the excellent
agreement with the analytical result (\ref{e17}).

\textit{General case.---} In the general case of a system with $\kappa$ components the second line of the Hamiltonian (\ref{Hc}) becomes $V(x,t)
\Psi^\dagger \Psi-\Psi^\dagger \boldsymbol{\mu}\Psi$ where now $ \Psi^\dagger=\left( \Psi_1^\dagger(x),\cdots,\Psi_\kappa^\dagger(x)\right)$ with
$\Psi_{\sigma}(x)$ $(\sigma=\{1,\cdots,\kappa\})$  fermionic or  bosonic fields satisfying the commutation  relations $\Psi_\sigma(x)\Psi_{\sigma'
}^\dagger(y)-\varepsilon\Psi_{\sigma'}^\dagger(y) \Psi_\sigma(x)=\delta_{\sigma\sigma'}\delta(x-y)$ and $\boldsymbol{\mu}$ is a diagonal matrix
with $(\mu_1,\cdots,\mu_\kappa)$ on the diagonal which are the chemical potentials of each component. The eigenstates of the system are described
by $\kappa$ sets of parameters \cite{S68,S93} $\boldsymbol{j}=\{j_i\}_{i=1}^N$ and $[\boldsymbol{\lambda}]=(\{ \lambda_i^{(1)}\}_{i=1}^{N_1},
\cdots,\{ \lambda_i^{(\kappa-1)}\}_{i=1}^{N_{\kappa-1}}) $ with $N\ge N_1\ge\cdots\ge N_{\kappa-1}\ge 0$ and will be denoted by $|\Phi^\kappa
(\boldsymbol{j},[\boldsymbol{\lambda}])\rangle$. The number of particles in the state $\sigma$ is $m_\sigma=N_{\sigma-1}-N_\sigma$ where we
consider $N_0=N$ and $N_\kappa=0$ and $\mathcal{H} |\Phi^\kappa(\boldsymbol{j},[\boldsymbol{\lambda}])\rangle=E_\kappa (\boldsymbol{j},[\boldsymbol
{\lambda}])|\Phi^\kappa(\boldsymbol{j},[\boldsymbol{\lambda}])\rangle$ with $|E_\kappa(\boldsymbol{j}, [\boldsymbol{\lambda}])\rangle=\sum_{i=1}^N
\hbar\omega_0(j_i+1/2)-\sum_{\sigma=1}^\kappa\mu_\sigma(N_{\sigma-1}-N_\sigma)$. The energies of  the eigenstates do not depend on the spin
configuration $[\boldsymbol{\lambda}]$  resulting in large degeneracies.  From now on we will consider the case of pure Zeeman splitting which is
described by $\mu_1=\mu-B(\kappa-1)$  and $\mu_{i+1}-\mu_i=2B$. The initial grandcanonical thermal state (analogue of (\ref{e10})) is
\begin{align*}
\boldsymbol{\rho}^{\mu,B,T}_\kappa=&\sum_{N=0}^\infty \sum_{N_1=0}^N\cdots\sum_{N_{\kappa-1}=0}^{N_{\kappa-2}}\sum_{\{\boldsymbol{j}\}}
\sum_{\{\boldsymbol{\lambda}^{(1)}\}}
\cdots \sum_{\{\boldsymbol{\lambda}^{(\kappa-1)}\}}\nonumber\\
& \times p_\kappa^{\boldsymbol{j},[\boldsymbol{\lambda}]}(\mu,B,T)|\Phi^\kappa(\boldsymbol{j},[\boldsymbol{\lambda}])\rangle
\langle\Phi^\kappa(\boldsymbol{j},[\boldsymbol{\lambda}])|\, ,\ \
\end{align*}
where now $p_\kappa^{\boldsymbol{j},[\boldsymbol{\lambda}]}(\mu,B,T)=e^{-E_\kappa(\boldsymbol{j},[\boldsymbol{\lambda}])/k_B T}/ \mathcal{Z}_{\kappa}
(\mu,B,T)$, with $\mathcal{Z}_{\kappa}(\mu,B,T)=\mbox{Tr}[e^{-\mathcal{H}^I_\kappa/k_B T}]$  the partition function of the system with $\kappa$
components at $t=0$. Like in the two-component case (see \cite{SM}) it can be shown that $\mathcal{Z}_{\kappa}(\mu,B,T)=Z_{\textsf{FF}}(\mu_\kappa'
,T)$  but now the renormalized chemical potential is (in the homogeneous case this result was first obtained by Schlottmann in \cite{S93})
\be\label{mug}
\mu_\kappa'=\mu+k_BT\ln[\sinh(\kappa B/k_BT)/\sinh(B/k_B T)]\, .
\ee

The calculations in the general case are very similar with the ones for the GY model. We now have $\kappa$ field correlators
$
g^{\mu,B,T}_\sigma(\xi_1,\xi_2;t)=\mbox{Tr}\left[\boldsymbol{\rho}^{\mu,B,T}_\kappa \Psi^\dagger_\sigma(\xi_1,t)\Psi_\sigma(\xi_2,t)\right]\, ,
$
$(\sigma=\{1,\cdots,\kappa\})$ and the same number of densities  $\rho^{\mu,B,T}_\sigma(\xi,t)= g^{\mu,B,T}_\sigma(\xi,\xi;t)$ and momentum distributions
$n^{\mu,B,T}_\sigma(p,t)$. Similar to the GY case the wavefunction  has a product  form with the  charge component given by a Slater determinant of
Hermite functions and the spin component given by an arbitrary function  of an appropriate spin chain \cite{DFBB08,GCWM09}. This  means that the mean
values of bilocal operators $G_{\sigma}^{\boldsymbol{j},[\boldsymbol{\lambda}]} (\xi_1,\xi_2)=\langle\Phi^\kappa(\boldsymbol{j},[ \boldsymbol{\lambda}])
|\Psi^\dagger_\sigma(\xi_1)\Psi_\sigma(\xi_2)|\Phi^\kappa (\boldsymbol{j},[\boldsymbol{\lambda}])\rangle$  appearing in the generalization of
Eq.~(\ref{e12}) also have a product representation generalizing (\ref{e13}) and given by (explicit expressions for the components can be found in
\cite{YGP15,DBS16,ASYP21,SM})
$
G_{\sigma}^{\boldsymbol{j},[\boldsymbol{\lambda}]} (\xi_1,\xi_2)=\sum_{d_1,d_2=1}^N S_\sigma(d_1,d_2) I(d_1,d_2;\xi_1,\xi_2)\, .
$
Unfortunately we do not know the value of $S_\sigma(d,d)$ (a reasonable conjecture would be $S_\sigma(d,d)=m_\sigma/N$) only that $\sum_{\sigma=1}^\kappa
S_\sigma(d,d)=1$ \cite{ASYP21}. Using this relation we obtain for the real space densities at $t=0$ $\sum_{\sigma=1}^\kappa \rho_\sigma^{\mu,B,T}(\xi)=
\rho_{\textsf{FF}}^{\mu_\kappa',T}(\xi)\, $ with $\mu_\kappa'$ defined in (\ref{mug}). In the large $t$ limit performing the stationary phase analysis
like in the GY case we obtain that the total asymptotic momentum distribution  has the same shape as the real space density profile $\sum_{\sigma=1}^\kappa
n_\sigma^{\mu,B,T}(p,t)\underset{t \rightarrow \infty} {\sim}  l_0^2\, \rho_{\textsf{FF}}^{\mu_{\kappa}',T}\left(p l_0^2/\hbar\right)$ and using
$n_{\textsf{FF}}^{\mu,T}(p)=l_0^2\, \rho_{\textsf{FF}}^{\mu,T}\left(p l_0^2/\hbar\right)$ we find
\be
\sum_{\sigma=1}^\kappa n_\sigma^{\mu,B,T}(p,t)\underset{t \rightarrow \infty} {\sim}n_{\textsf{FF}}^{\mu_\kappa',T}\left(p\right)\, ,
\ee
which is the dynamical fermionization of the strongly interacting $\kappa$ component gas.

\textit{Finite interaction case.---} In the case of large, but finite repulsion, we expect that most of the features presented above to remain
valid \cite{ASYP21}. In this case, to first order in $g$, the wavefunctions still have a product form \cite{YGP15} with the charge degrees of
freedom characterized by a Slater determinant and the spin part described by a spin chain [antiferromagnetic (ferromagnetic) in the fermionic
(bosonic) case] with position dependent coefficients $C_i$. Fortunately, the time-evolution of these coefficients during expansion is given by
$C_i(t)=b^{-3}(t)C_i(0)$  \cite{VHZ16} which means that spin dynamics of the  system remains frozen like in the impenetrable case and the same
considerations apply.
For  arbitrary repulsion it is also sensible to assume that the system will dynamically fermionize after expansion and that the initial quasimomenta
of the trapped gas will be mapped to real momenta of the expanded cloud similar to the case of single component bosons \cite{JPGB08,IA12,CGK15}.
This is due to the fact that at long time after release  the dimensionless parameter $\gamma(x)=c/n(x)$ ($c= m g/\hbar^2$), which characterizes the
strength of the interaction, will  become very large (the density $n(x)$ decreases) and, therefore, the  dynamics will be described by the TG
Hamiltonian [(\ref{Hc}) with $g=\infty$]. We expect that these considerations can be made rigorous using the Yudson representation for integrable
systems \cite{Y85} generalizing the proof  for the  Lieb-Liniger model  derived in \cite{IA12}.

\textit{Conclusions.---}
We have proved that DF occurs in all bosonic and fermionic impenetrable 1D spinor gases at finite temperature. At long times after release from
the trap the asymptotic momentum distribution approaches that of a system of spinless noninteracting fermions at the same temperature and a renormalized
chemical potential which depends on the number of the components of the spinor system and magnetic field but not on the statistics.  Using the same method
one can prove the existence of DF  in the case of an arbitrary Bose-Fermi mixture \cite{S68,ID06,P22,FVGM11,DJAR17} using the fact that the wavefunctions in the TG regime also factorize
with the spin component given by wavefunctions of an appropriate graded spin-chain while the charge part is still described by a  Slater determinant of
Hermite functions. The proof runs along the same lines taking into account that the thermodynamics (partition function) of impenetrable particles is
independent of the statistics of the constituent particles.

\acknowledgments
Financial support  from the Grants No. 16N/2019 and 30N/2023 of the National Core Program  of the Romanian Ministry of Research, Innovation and Digitization is gratefully
acknowledged.

\bigskip

\onecolumngrid
\newpage



\section*{Supplementary material}

# Supplemental Material: Dynamical fermionization in one-dimensional spinor gases at finite temperature

Ovidiu I. Pâţu[1]

[1]*Institute for Space Sciences, Bucharest-Măgurele, R 077125, Romania*

## I. DYNAMICAL FERMIONIZATION IN THE LIEB-LINIGER MODEL AT FINITE TEMPERATURE

Here we prove analytically that dynamical fermionization occurs at finite temperature in the Lieb-Liniger model. The Lieb-Liniger model describes bosons with repulsive contact interactions in one dimension. In the presence of an external time-dependent harmonic potential $V(x,t) = \theta(-t)m\omega_0^2 x^2/2$ the Hamiltonian in second quantization is

$$\mathcal{H}_{LL} = \int dx \, \frac{\hbar^2}{2m} \left(\partial_x \Psi^\dagger\right)(\partial_x \Psi) + g\Psi^\dagger \Psi^\dagger \Psi\Psi + (V(x,t) - \mu)\Psi^\dagger \Psi \,, \tag{S-1}$$

where $m$ is the mass of the particles and $g > 0$ the strength of the repulsive interaction. In (S-1) $\Psi^\dagger(x)$ and $\Psi(x)$ are bosonic fields satisfying the canonical commutation relations $\Psi(x)\Psi^\dagger(y) = \Psi^\dagger(y)\Psi(x) + \delta(x-y)$. The homogeneous Lieb-Liniger model is integrable for any value of the coupling strength. In the presence of the trapping potential we have an analytical solution only for $g=0$ and in the case of hardcore particles ($g=\infty$) which will be investigated below. At $t=0$ and $g=\infty$ the eigenstates of the trapped Lieb-Liniger model are

$$|\Phi_N(\boldsymbol{j})\rangle = \frac{1}{\sqrt{N!}} \int dx \, \chi_N(x_1,\cdots,x_N|\boldsymbol{j}) \Psi^\dagger(x_N) \cdots \Psi^\dagger(x_1)|0\rangle \,, \tag{S-2}$$

with $|0\rangle$ is the Fock vacuum satisfying $\Psi(x)|0\rangle = 0$ for all $x$. The wavefunctions are obtained from the Bose-Fermi mapping as

$$\chi_N(x_1,\cdots,x_N|\boldsymbol{j}) = \prod_{j>k} \text{sign}(x_j - x_k) \chi_N^{\mathsf{FF}}(x_1,\cdots,x_N|\boldsymbol{j}) \,, \tag{S-3}$$

with $\boldsymbol{j} = (j_1,\cdots,j_N)$ describing the single particle energies and

$$\chi_N^{\mathsf{FF}}(x_1,\cdots,x_N|\boldsymbol{j}) = \frac{1}{\sqrt{N!}} \det_N \left[\phi_{j_a}(x_b)\right] \,, \tag{S-4}$$

is the wavefunction of a dual system of free spinless fermions in the presence of the same trapping potential. Here $\phi_j(x)$ are the quantum harmonic oscillator wavefunctions of frequency $\omega_0$ defined by $\phi_j(x) = \left(2^j j!\right)^{-1/2} \left(\frac{m\omega_0}{\pi\hbar}\right)^{1/4} e^{-\frac{m\omega_0 x^2}{2\hbar}} H_j\left(\sqrt{\frac{m\omega_0}{\hbar}}x\right)$ with $H_j(x)$ the Hermite polynomials. The eigenstates (S-2) are normalized $\langle \Phi_N(\boldsymbol{j})|\Phi_N(\boldsymbol{j}')\rangle = \delta_{\boldsymbol{j},\boldsymbol{j}'}$ and $\mathcal{H}_{LL}|\Phi_N(\boldsymbol{j})\rangle = E_N(\boldsymbol{j})|\Phi_N(\boldsymbol{j})\rangle$ with $E_N(\boldsymbol{j}) = \sum_{i=1}^N \left[\hbar\omega_0(j_i + 1/2) - \mu\right]$.

We are interested in the time evolution of the field-field correlator (one-body reduced density matrix) at finite temperature after release from the trap. We consider the system initially prepared in a finite temperature state characterized by the grandcanonical density matrix

$$\boldsymbol{\rho}^{\mu,T} = \sum_{N=0}^\infty \sum_{\{\boldsymbol{j}\}} p_N^{\boldsymbol{j}}(\mu,T) |\Phi_N(\boldsymbol{j})\rangle\langle\Phi_N(\boldsymbol{j})|, \quad p_N^{\boldsymbol{j}}(\mu,T) = \exp\left[-E_N(\boldsymbol{j})/k_B T\right]/\mathcal{Z}(\mu,T) \,, \tag{S-5}$$

where $\mathcal{Z}(\mu,T) = \text{Tr}[e^{-\mathcal{H}_{LL}^I/k_B T}] = \sum_{N=0}^\infty \sum_{\{\boldsymbol{j}\}} e^{-E_N(\boldsymbol{j})/k_B T}$ with $\mathcal{H}_{LL}^I$ the Hamiltonian (S-1) at $t=0$. At $t>0$ we release the system from the trap and we let it evolve with $\mathcal{H}_{LL}^F$ which is given by (S-1) with $V(x,t) = 0$. The main object of interest for us is the field-field correlator

$$g^{\mu,T}(\xi_1,\xi_2;t) = \text{Tr}\left[\boldsymbol{\rho}^{\mu,T} \Psi^\dagger(\xi_1,t)\Psi(\xi_2,t)\right] \,, \tag{S-6}$$

with $\Psi^\dagger(\xi,t) = e^{i\mathcal{H}_{LL}^F t}\Psi^\dagger(\xi)e^{-i\mathcal{H}_{LL}^F t}$. From the correlator one can obtain the real space density $\rho^{\mu,T}(\xi,t) = g^{\mu,T}(\xi,\xi;t)$ and the momentum distribution $n^{\mu,T}(p,t) = \int e^{ip(\xi_1-\xi_2)/\hbar} g^{\mu,T}(\xi_1,\xi_2;t) \, d\xi_1 d\xi_2/2\pi$.

The field-field correlator can be written as

$$g^{\mu,T}(\xi_1,\xi_2;t) = \sum_{N=1}^{\infty} \sum_{\{\boldsymbol{j}\}} p_N^{\boldsymbol{j}}(\mu,T)\, G_N^{\boldsymbol{j}}(\xi_1,\xi_2;t)\,, \qquad (\text{S-7})$$

with

$$G_N^{\boldsymbol{j}}(\xi_1,\xi_2;t) = N \int \prod_{k=2}^{N} dx_k\, \bar{\chi}_N(\xi_1,x_2,\cdots,x_N;t|\boldsymbol{j})\chi_N(\xi_2,x_2,\cdots,x_N;t|\boldsymbol{j})\,, \qquad (\text{S-8})$$

where $\chi_N(x_1,x_2,\cdots,x_N;t|\boldsymbol{j})$ are the wavefunctions (S-3) time-evolved with $\mathcal{H}_{LL}^F$. The wavefunctions (S-3) are expressed in terms of the Hermite functions and the time evolution of the harmonic oscillator with variable frequency is given by the scaling transformation [S1, S2] $\phi_j(x,t) = \frac{1}{\sqrt{b}}\phi_j\left(\frac{x}{b},0\right)\exp\left[i\frac{mx^2}{2\hbar}\frac{\dot{b}}{b} - iE_j\tau(t)\right]$ where $b(t)$ is a solution of the Ermakov-Pinney equation $\ddot{b} = -\omega(t)^2 b + \omega_0^2/b^3$ with boundary conditions $b(0)=1$, $\dot{b}(0)=0$, $E_j = \hbar\omega_0(j+1/2)$ and the rescaled time parameter is given by $\tau(t) = \int_0^t dt'/b^2(t')$. This means that the dynamics of the $G_N^{\boldsymbol{j}}$ functions is given by ($l_0 = \sqrt{\hbar/(m\omega_0)}$)

$$G_N^{\boldsymbol{j}}(\xi_1,\xi_2;t) = \frac{1}{b} G_N^{\boldsymbol{j}}\left(\frac{\xi_1}{b},\frac{\xi_2}{b};0\right) e^{-\frac{i}{b}\frac{\dot{b}}{\omega_0}\frac{\xi_1^2-\xi_2^2}{2l_0^2}}\,, \qquad (\text{S-9})$$

and introducing the notation $\widetilde{G}_N^{\boldsymbol{j}}(p,t) = \int e^{ip(\xi_1-\xi_2)/\hbar} G_N^{\boldsymbol{j}}(\xi_1,\xi_2;t)\, d\xi_1 d\xi_2/2\pi$ we have

$$\widetilde{G}_N^{\boldsymbol{j}}(p,t) = \frac{b}{2\pi}\int G_N^{\boldsymbol{j}}(\xi_1,\xi_2;0) e^{-ib\left[\frac{\dot{b}}{\omega_0}\frac{\xi_1^2-\xi_2^2}{2l_0^2} - \frac{p(\xi_1-\xi_2)}{\hbar}\right]} d\xi_1 d\xi_2\,. \qquad (\text{S-10})$$

Now we can look at dynamical fermionization. The first step is to investigate the density at $t=0$. From (S-7), (S-8) and (S-3) we obtain $\rho^{\mu,T}(\xi) = \sum_{N=1}^{\infty}\sum_{\{\boldsymbol{j}\}} p_N^{\boldsymbol{j}}(\mu,T) G_N^{\boldsymbol{j}}(\xi,\xi)$. Noticing that $G_N^{\boldsymbol{j}}(\xi,\xi)$ for the TG gas is the same as the similar expression for free fermions $G_{N,\mathsf{FF}}^{\boldsymbol{j}}(\xi,\xi)$ defined as in (S-8) but with $\chi_N^{\mathsf{FF}}$ for the wavefunctions and that the partition functions of free fermions and the TG gas are the same this proves that the initial density is the same as the density of free fermions at the same temperature and chemical potential $\rho^{\mu,T}(\xi) = \rho_{\mathsf{FF}}^{\mu,T}(\xi)$. Let's investigate the dynamics. In the case of free expansion the solution of the Ermakov-Pinney equation is $b(t) = (1+\omega_0^2 t^2)^{1/2}$ and in the large time limit we have $\lim_{t\to\infty} b(t) = \omega_0 t$ and $\lim_{t\to\infty} \dot{b}(t) = \omega_0$. In the same limit we have $\lim_{t\to\infty} n(p,t) = \sum_{N=1}^{\infty}\sum_{\{\boldsymbol{j}\}} p_N^{\boldsymbol{j}}(\mu,T)\lim_{t\to\infty}\widetilde{G}_N^{\boldsymbol{j}}(p,t)$. Using the method of stationary phase in (S-10) (the points of stationary phase are $\xi_0 = p\omega_0 l_0^2/(\dot{b}\hbar)$ for both integrals) we find $\widetilde{G}_N^{\boldsymbol{j}}(p,t) \underset{t\to\infty}{\sim} \left|\frac{\omega_0 l_0^2}{\dot{b}}\right| G_N^{\boldsymbol{j}}\left(\frac{p\omega_0 l_0^2}{\dot{b}\hbar},\frac{p\omega_0 l_0^2}{\dot{b}\hbar};0\right)$. But $G_N^{\boldsymbol{j}}(p,p;0) = G_{N,\mathsf{FF}}^{\boldsymbol{j}}(p,p;0)$ which means that

$$\lim_{t\to\infty} n(p,t) = \sum_{N=1}^{\infty}\sum_{\{\boldsymbol{j}\}} p_N^{\boldsymbol{j}}(\mu,T) \left|\frac{\omega_0 l_0^2}{\dot{b}}\right| G_{N,\mathsf{FF}}^{\boldsymbol{j}}\left(\frac{p\omega_0 l_0^2}{\dot{b}\hbar},\frac{p\omega_0 l_0^2}{\dot{b}\hbar};0\right)\,, \qquad (\text{S-11})$$

or $n(p,t) \underset{t\to\infty}{\sim} l_0^2 \rho_{\mathsf{FF}}^{\mu,T}\left(pl_0^2/\hbar,0\right)$ which shows that the momentum distribution has the same shape as the initial density. Using the identity $n_{\mathsf{FF}}^{\mu,T}(p) = l_0^2\, \rho_{\mathsf{FF}}^{\mu,T}\left(pl_0^2/\hbar\right)$ where $n_{\mathsf{FF}}^{\mu,T}(p)$ is the momentum distribution of spinless noninteracting fermions at temperature $T$ and chemical potential $\mu$ we find that in the large $t$ limit

$$n^{\mu,T}(p,t) \underset{t\to\infty}{\sim} n_{\mathsf{FF}}^{\mu,T}(p)\,, \qquad (\text{S-12})$$

proving the existence of dynamical fermionization in the Lieb-Liniger model at finite temperature.

## II. NEW PARAMETRIZATION OF THE CORRELATORS IN THE GAUDIN-YANG MODEL

In this section we are going to introduce a new parametrization of the wavefunctions which will make explicit the factorization (7) of the mean value of the bilocal operators. Using the commutation relations and the explicit expression of the eigenstates (2) it can be shown that $G_{N,M,\sigma}^{\boldsymbol{j},\boldsymbol{\lambda}}(\xi_1,\xi_2) = \langle\Phi_{N,M}(\boldsymbol{j},\boldsymbol{\lambda})|\Psi_\sigma^\dagger(\xi_1)\Psi_\sigma(\xi_2)|\Phi_{N,M}(\boldsymbol{j},\boldsymbol{\lambda})\rangle$ can be expressed as

$$G^{\boldsymbol{j},\boldsymbol{\lambda}}_{N,M,\downarrow}(\xi_1,\xi_2) = \frac{(N!)^2}{c_\downarrow} \int \prod_{j=2}^{N} dx_j \, \bar{\chi}^{\boldsymbol{\alpha}}_{N,M}(\xi_1,x_2,\cdots,x_N|\boldsymbol{j},\boldsymbol{\lambda}) \chi^{\boldsymbol{\alpha}}_{N,M}(\xi_2,x_2,\cdots,x_N|\boldsymbol{j},\boldsymbol{\lambda}),\tag{S-13a}$$

$$G^{\boldsymbol{j},\boldsymbol{\lambda}}_{N,M,\uparrow}(\xi_1,\xi_2) = \frac{(N!)^2}{c_\uparrow} \int \prod_{j=1}^{N-1} dx_j \, \bar{\chi}^{\boldsymbol{\alpha}}_{N,M}(x_1,\cdots,x_{N-1},\xi_1|\boldsymbol{j},\boldsymbol{\lambda}) \chi^{\boldsymbol{\alpha}}_{N,M}(x_1,\cdots,x_{N-1},\xi_2|\boldsymbol{j},\boldsymbol{\lambda}),\tag{S-13b}$$

where $\boldsymbol{\alpha} = (\downarrow\cdots\downarrow\uparrow\cdots\uparrow)$, which is called the canonical ordering, $c_\downarrow = (N-M)!(M-1)!$ and $c_\uparrow = (N-M-1)!M!$. The time evolution of the $G$ functions described in Eq. (8) of the main text can be easily derived by plugging the scaling transformation of the Hermite functions in (S-13).

We will use a new parametrization of the wavefunction, originally introduced in the investigation of the correlation functions in the Bose-Fermi mixture [S3, S4], which makes explicit the spin and charge factorization in (S-13). In the canonical ordering $x_1,\cdots,x_M$ denote the coordinates of the spin down particles while $x_{M+1},\cdots,x_N$ the coordinates of the spin up particles. We introduce a new set of ordered coordinates $\boldsymbol{z} = (z_1,\cdots,z_N)$ defined in the wedge

$$Z = \{-\infty \leq z_1 \leq z_2 \leq \cdots \leq z_N \leq +\infty\},\tag{S-14}$$

which describe the positions of the particles independent of their spin. For a given configuration $\boldsymbol{x} = (x_1,\cdots,x_N)$ all the sets $P\boldsymbol{x} = (x_{P_1},\cdots,x_{P_N})$ with $P \in S_N$ an arbitrary permutation are described by the same set of $(z_1,\cdots,z_N)$. We also need to introduce a new set of variables $\boldsymbol{y} = (y_1,\cdots,y_N)$ which specify the position of the particles in the ordered set $(z_1,\cdots,z_N)$. The subset $\{y_1,\cdots,y_M\}$ describe the position of the spin up particles while $\{y_{M+1},\cdots,y_N\}$ are the positions of the spin down particles and we have $z_{y_i} = x_i$. For example in the case of a system in the $(N,M) = (6,2)$-sector (remember we are in the canonical ordering which means that the first two particles have spin down) with $x_3 < x_1 < x_2 < x_4 < x_5 < x_6$ then $\boldsymbol{z} \equiv (z_1,z_2,z_3,z_4,z_5,z_6) = (x_3,x_1,x_2,x_4,x_5,x_6)$ with $\boldsymbol{y} = (2,3,1,4,5,6)$. Note that $y_1,\cdots,y_M$ are the same as $n_1,\cdots,n_M$ in the original definition of the wavefunction. Because $(-1)^P = \prod_{N\geq j>k\geq 1}\text{sign}(y_j - y_k)$ the wavefunctions in the new parametrization are (we drop the superscript because from now on we will always use the canonical ordering $\boldsymbol{\alpha} = (\downarrow\cdots\downarrow\uparrow\cdots\uparrow)$)

$$\chi_{N,M}(\boldsymbol{z},\boldsymbol{y}|\boldsymbol{j},\boldsymbol{\lambda}) = \frac{1}{N!N^{M/2}} h_\varepsilon(\boldsymbol{y}) \det_M\left(e^{iy_a\lambda_b}\right) \det_N\left[\phi_{j_a}(z_b)\right],\tag{S-15}$$

where

$$h_\varepsilon(\boldsymbol{y}) = \begin{cases} \prod_{N\geq j>k\geq M+1} \text{sign}(y_j - y_k) & \varepsilon = 1 \quad \text{(bosons)}, \\ \prod_{M\geq j>k\geq 1} \text{sign}(y_j - y_k) & \varepsilon = -1 \quad \text{(fermions)}. \end{cases}\tag{S-16}$$

In the new parametrization the $G$ functions for $\xi_1 \leq \xi_2$ can be written in a factorized form (the computations are similar with the ones presented in Appendix C of [S4] for the Bose-Fermi mixture)

$$G^{\boldsymbol{j},\boldsymbol{\lambda}}_{N,M,\sigma}(\xi_1,\xi_2) = \frac{1}{c_\sigma N^M} \sum_{d_1=1}^{N} \sum_{d_2=d_1}^{N} S_\sigma(d_1,d_2) I(d_1,d_2;\xi_1,\xi_2),\tag{S-17}$$

with the charge functions $I(d_1,d_2;\xi_1,\xi_2)$ having the same expressions for both spin up and spin down cases

$$I(d_1,d_2;\xi_1,\xi_2) = \int_{Z_{d_1,d_2}(\xi_1,\xi_2)} \prod_{j=1,j\neq d_1}^{N} dz_j \det_N\left[\bar{\phi}_{j_a}(z_b)\right] \det_N\left[\phi_{j_a}(z'_b)\right],\tag{S-18}$$

with $Z_{d_1,d_2}(\xi_1,\xi_2)$ is defined as

$$Z_{d_1,d_2}(\xi_1,\xi_2) = \{-\infty \leq z_1 \leq \cdots \leq z_{d_1-1} \leq \xi_1 \leq z_{d_1+1} \leq \cdots \leq z_{d_2} \leq \xi_2 \leq z_{d_2+1} \leq \cdots \leq z_N \leq +\infty\},\tag{S-19}$$

and $\boldsymbol{z}'$ satisfies the constraint

$$-\infty \leq z'_1 = z_1 \leq \cdots \leq z'_{d_1-1} = z_{d_1-1} \leq \xi_1 \leq z'_{d_1} = z_{d_1+1} \leq \cdots$$
$$\cdots \leq z'_{d_2-1} = z_{d_2} \leq \xi_2 \leq z'_{d_2+1} = z_{d_2+1} \leq \cdots \leq z'_N = z_N \leq +\infty.\tag{S-20}$$

The spin functions are different. For spin down particles we have

$$S_\downarrow(d_1,d_2) = \sum_{\boldsymbol{y}\in Y(y_1=d_1)} h_\varepsilon(\boldsymbol{y}) h_\varepsilon(\boldsymbol{y}') \det_M\left(e^{-iy_a\lambda_b}\right) \det_M\left(e^{iy'_a\lambda_b}\right).\tag{S-21}$$

with $Y(y_1 = d) = \{\boldsymbol{y} \in S_N | \text{ with } y_1 = d\}$ and the connection between $\boldsymbol{y}$ and $\boldsymbol{y}'$ is given by

$$\begin{aligned} y_1' &= d_2\,, \quad y_1 = d_1\,, \\ y_i' &= y_i \quad \text{for } y_i < d_1\,, \\ y_i' &= y_i - 1 \quad \text{for } d_1 < y_i \leq d_2\,, \\ y_i' &= y_i\,, \quad \text{for } d_2 < y_i\,. \end{aligned} \tag{S-22}$$

while for spin up particles we have

$$S_\uparrow(d_1, d_2) = \sum_{\boldsymbol{y} \in Y(y_N = d_1)} h_\varepsilon(\boldsymbol{y}) h_\varepsilon(\boldsymbol{y}') \det_M\left(e^{-iy_a \lambda_b}\right) \det_M\left(e^{iy_a' \lambda_b}\right), \tag{S-23}$$

with the connection between $\boldsymbol{y}$ and $\boldsymbol{y}'$ given by ($y_1, \cdots, y_M$ and $y_1', \cdots, y_M'$ cannot take the values $d_1$ and $d_2$)

$$\begin{aligned} y_N' &= d_2\,, \quad y_N = d_1\,, \\ y_i' &= y_i \quad \text{for } y_i < d_1\,, \\ y_i' &= y_i - 1 \quad \text{for } d_1 < y_i \leq d_2\,, \\ y_i' &= y_i\,, \quad \text{for } d_2 < y_i\,. \end{aligned} \tag{S-24}$$

In the region $\xi_2 < \xi_1$ the mean values can be obtained via complex conjugation $G_{N,M,\sigma}^{\boldsymbol{j},\boldsymbol{\lambda}}(\xi_1, \xi_2) = \overline{G_{N,M,\sigma}^{\boldsymbol{j},\boldsymbol{\lambda}}(\xi_2, \xi_1)}$.

In fact for the analytical derivation of the dynamical fermionization we need to evaluate the $G$ functions in the simpler case of $\xi_1 = \xi_2 = \xi$ for which

$$G_{N,M,\sigma}^{\boldsymbol{j},\boldsymbol{\lambda}}(\xi, \xi) = \frac{1}{c_\sigma N^M} \sum_{d=1}^N S_\sigma(d,d) I(d,d;\xi,\xi)\,, \tag{S-25}$$

with

$$I(d,d;\xi,\xi) = \int_{Z_d(\xi)} \prod_{j=1, j\neq d}^N dz_j \det_N\left[\bar\phi_{j_a}(z_b)\right] \det_N\left[\phi_{j_a}(z_b)\right], \tag{S-26}$$

with $Z_d(\xi) = \{-\infty \leq z_1 \leq \cdots \leq z_{d-1} \leq \xi \leq z_{d+1} \leq \cdots \leq z_N \leq +\infty\}$. When $d_1 = d_2 = d$ the relation between $\boldsymbol{y}$ and $\boldsymbol{y}'$ for the spin down function $S_\downarrow(d,d)$ is $\boldsymbol{y} = (d, y_2, \cdots, y_N)\,, \boldsymbol{y}' = (d, y_2, \cdots, y_N)$ and for the spin up function $S_\downarrow(d,d)$ we have $\boldsymbol{y} = (y_1, \cdots, y_{N-1}, d)\,, \boldsymbol{y}' = (y_1, \cdots, y_{N-1}, d)$ with the restriction that $y_1, \cdots, y_M$ cannot take the values $d$. Doing similar calculations like in Appendix D of [S4] we obtain

$$S_\downarrow(d,d) = (N-M)! M! N^{M-1}\,, \quad S_\uparrow(d,d) = (N-M-1)! M! N^{M-1}(N-M) \text{ for any } d\,, \tag{S-27}$$

and introducing new variables $t_1, \cdots, t_{N-1}$

$$\sum_{d=1}^N I(d,d;\xi,\xi) = N! \int_{t_1 \leq \cdots \leq t_{N-1}} \overline{\chi}_N^{\mathsf{FF}}(\xi, t_1, \cdots, t_{N-1}) \chi_N^{\mathsf{FF}}(\xi, t_1, \cdots, t_{N-1})\, dt_1 \cdots dt_{N-1}\,, \tag{S-28}$$

with $\chi_N^{\mathsf{FF}}(\boldsymbol{x}|\boldsymbol{j}) = \det_N[\phi_{j_a}(x_b)]/\sqrt{N!}$ the wavefunction of free fermions. Using the symmetry of the integrand we obtain $\sum_{d=1}^N I(d,d;\xi,\xi) = N \int_{-\infty}^{+\infty} \prod_{j=1}^{N-1} dt_j\, \overline\chi_N^{\mathsf{FF}}(\xi, t_1, \cdots, t_{N-1}) \chi_N^{\mathsf{FF}}(\xi, t_1, \cdots, t_{N-1})$ which is the density of free fermions (in the state $\boldsymbol{j}$) at position $\xi$.

## III. ANALYTICAL DERIVATION OF DYNAMICAL FERMIONIZATION IN THE GAUDIN-YANG MODEL

Due to space limitations in the main text we have only outlined the derivation of dynamical fermionization in the Gaudin-Yang model. Here we provide the full details. We start with the derivation of the partition function.

*Partition function of the impenetrable Gaudin-Yang model.—* At $t = 0$ the energy of an eigenstate of the trapped, bosonic or fermionic, impenetrable Gaudin-Yang model with $N$ particles of which $M$ have spin-down is given by

$$E_{N,M}(\boldsymbol{j}, \boldsymbol{\lambda}) = \sum_{i=1}^N [\hbar\omega_0(j_i + 1/2) - \mu + B] - 2BM. \tag{S-29}$$

We point out two important features of the spectrum: a) it is independent of the statistics of the particles (this is of course valid only for impenetrable particles) and b) does not depend on the spin state $\boldsymbol{\lambda}$ and, therefore, it is highly degenerate. In the $(N, M)$-sector the spin states are given by $\boldsymbol{\lambda} = (\lambda_1, \cdots, \lambda_M)$ with $e^{i\lambda_a N} = (-1)^{M+1}$. The $\lambda$'s can take $N$ values which means that there are $C_N^M$ spin states which is also the degeneracy of an arbitrary eigenstate described by $\boldsymbol{j}$. The partition function of the system at $=0$ is given by

$$\mathcal{Z}(\mu, B, T) = \text{Tr}[e^{-\mathcal{H}^I/k_B T}],$$

$$= \sum_{N=0}^{\infty} \sum_{M=0}^{N} \sum_{j_1 < \cdots < j_N} \sum_{\lambda_1 < \cdots < \lambda_M} e^{-E_{N,M}(\boldsymbol{j}, \boldsymbol{\lambda})/k_B T},$$

$$= \sum_{N=0}^{\infty} \sum_{M=0}^{N} \sum_{j_1 < \cdots < j_N} \sum_{\lambda_1 < \cdots < \lambda_M} e^{2BM/k_B T} e^{-\sum_{i=1}^{N}[\hbar\omega_0(j_i+1/2)-\mu+B]/k_B T},$$

$$= \sum_{N=0}^{\infty} \sum_{j_1 < \cdots < j_N} \left(1 + e^{2B/k_B T}\right)^N e^{-\sum_{i=1}^{N}[\hbar\omega_0(j_i+1/2)-\mu+B]/k_B T},$$

$$= \sum_{N=0}^{\infty} \sum_{j_1 < \cdots < j_N} e^{-\sum_{i=1}^{N}[\hbar\omega_0(j_i+1/2)-\mu']/k_B T}, \qquad (\text{S-30})$$

where we have used $\sum_{M=0}^{N} \sum_{\lambda_1 < \cdots < \lambda_M} e^{2BM/T} = \sum_{M=0}^{N} C_N^M e^{2BM/k_B T} = (1 + e^{2B/k_B T})^N$ and

$$\mu' = \mu + k_B T \ln[2\cosh(B/k_B T)]. \qquad (\text{S-31})$$

From (S-30) we see that the partition function of the harmonically trapped GY model in the TG regime is the same as the one of trapped spinless free fermions $\mathcal{Z}_{\text{FF}}(\mu', T)$ at the same temperature but with renormalized chemical potential given by (S-31).

*Densities of the trapped system.—* We now investigate the densities at $t = 0$. From the definition (Eq. (6) of the main text) we have $(p_{N,M}^{\boldsymbol{j}, \boldsymbol{\lambda}}(\mu, B, T) = e^{-E_{N,M}(\boldsymbol{j}, \boldsymbol{\lambda})/k_B T}/\mathcal{Z}(\mu, B, T))$

$$\rho_\sigma^{\mu, B, T}(\xi) = \sum_{N=1}^{\infty} \sum_{M=0}^{N} \sum_{\{\boldsymbol{j}\}} \sum_{\{\boldsymbol{\lambda}\}} p_{N,M}^{\boldsymbol{j}, \boldsymbol{\lambda}} G_{N,M,\sigma}^{\boldsymbol{j}, \boldsymbol{\lambda}}(\xi, \xi), \qquad (\text{S-32})$$

where $c_\downarrow = (N-M)!(M-1)!$ and $c_\uparrow = (N-M-1)!M!$ and (see (S-25))

$$G_{N,M,\sigma}^{\boldsymbol{j}, \boldsymbol{\lambda}}(\xi, \xi) = \sum_{d=1}^{N} S_\sigma(d, d) I(d, d; \xi, \xi)/(c_\sigma N^M). \qquad (\text{S-33})$$

$I(d, d; \xi, \xi)$ is defined in (S-26). From the previous section [see (S-27) and the discussion after (S-28)] for any value of $d$ (S-27) we have $S_\downarrow(d, d) = (N-M)!M! N^{M-1}$, $S_\uparrow(d, d) = (N-M-1)!M! N^{M-1}(N-M)$ and

$$\sum_{d=1}^{N} I(d, d; \xi, \xi) \equiv G_{N,\text{FF}}^{\boldsymbol{j}}(\xi, \xi) = N \int_{-\infty}^{+\infty} \prod_{j=1}^{N-1} dt_j \, \overline{\chi}_N^{\text{FF}}(\xi, t_1, \cdots, t_{N-1}) \chi_N^{\text{FF}}(\xi, t_1, \cdots, t_{N-1}), \qquad (\text{S-34})$$

which is the density of free fermions (in the state $\boldsymbol{j}$) at position $\xi$. Using (S-33) and (S-34) we obtain

$$\rho_\downarrow^{\mu, B, T}(\xi) = \sum_{N=1}^{\infty} \sum_{M=0}^{N} \sum_{\{\boldsymbol{j}\}} \sum_{\{\boldsymbol{\lambda}\}} \frac{M}{N} p_{N,M}^{\boldsymbol{j}, \boldsymbol{\lambda}} G_{N,\text{FF}}^{\boldsymbol{j}}(\xi, \xi), \qquad (\text{S-35})$$

$$\rho_\uparrow^{\mu, B, T}(\xi) = \sum_{N=1}^{\infty} \sum_{M=0}^{N} \sum_{\{\boldsymbol{j}\}} \sum_{\{\boldsymbol{\lambda}\}} \frac{N-M}{N} p_{N,M}^{\boldsymbol{j}, \boldsymbol{\lambda}} G_{N,\text{FF}}^{\boldsymbol{j}}(\xi, \xi). \qquad (\text{S-36})$$

Making use of $\mathcal{Z}(\mu, B, T) = \mathcal{Z}_{\text{FF}}(\mu', T)$ and of the identity

$$\sum_{M=0}^{N} \sum_{\{\boldsymbol{\lambda}\}} \frac{M}{N} e^{2BM/k_B T} = \frac{k_B T}{2N} \frac{\partial}{\partial B} \left( \sum_{M=0}^{N} \sum_{\{\boldsymbol{\lambda}\}} e^{2BM/k_B T} \right) = \frac{e^{B/k_B T}}{2\cosh(B/k_B T)} \left(1 + e^{2B/k_B T}\right)^N, \qquad (\text{S-37})$$

the density of spin down particles can be written as

$$\rho_\downarrow^{\mu,B,T}(\xi) = \sum_{N=1}^\infty \sum_{M=0}^N \sum_{\{j\}} \sum_{\{\lambda\}} \frac{M}{N} e^{2BM/k_BT} \frac{e^{-NB/k_BT}e^{-\sum_{i=1}^N[\hbar\omega_0(j_i+1/2)-\mu]/k_BT}}{\mathcal{Z}_{\mathsf{FF}}(\mu',T)} G_{N,\mathsf{FF}}^{\boldsymbol{j}}(\xi,\xi) ,$$

$$= \frac{e^{B/k_BT}}{2\cosh(B/k_BT)} \sum_{N=1}^\infty \sum_{\{j\}} \left(1+e^{2B/k_BT}\right)^N \frac{e^{-NB/k_BT}e^{-\sum_{i=1}^N[\hbar\omega_0(j_i+1/2)-\mu]/k_BT}}{\mathcal{Z}_{\mathsf{FF}}(\mu',T)} G_{N,\mathsf{FF}}^{\boldsymbol{j}}(\xi,\xi) ,$$

$$= \frac{e^{B/k_BT}}{2\cosh(B/k_BT)} \underbrace{\sum_{N=1}^\infty \sum_{\{j\}} \frac{e^{-\sum_{i=1}^N[\hbar\omega_0(j_i+1/2)-\mu']/k_BT}}{\mathcal{Z}_{\mathsf{FF}}(\mu',T)} G_{N,\mathsf{FF}}^{\boldsymbol{j}}(\xi,\xi)}_{\rho_{\mathsf{FF}}^{\mu',T}(\xi)} ,$$

(S-38)

which shows that

$$\rho_\downarrow^{\mu,B,T}(\xi) = \frac{e^{B/k_BT}}{2\cosh(B/k_BT)} \rho_{\mathsf{FF}}^{\mu',T}(\xi) . \tag{S-39}$$

In a similar fashion, using the identity

$$\sum_{M=0}^N \sum_{\{\lambda\}} \frac{N-M}{N} e^{2BM/k_BT} = \left(1+e^{2B/k_BT}\right)^N - \frac{e^{B/k_BT}}{2\cosh(B/k_BT)} \left(1+e^{2B/k_BT}\right)^N , \tag{S-40}$$

we can show that

$$\rho_\uparrow^{\mu,B,T}(\xi) = \frac{e^{-B/k_BT}}{2\cosh(B/k_BT)} \rho_{\mathsf{FF}}^{\mu',T}(\xi) . \tag{S-41}$$

From Eqs. (S-39) and (S-48) we have

$$\rho_\downarrow^{\mu,B,T}(\xi) + \rho_\uparrow^{\mu,B,T}(\xi) = \rho_{\mathsf{FF}}^{\mu',T}(\xi) , \tag{S-42}$$

proving that the initial densities are proportional to the densities of trapped spinless free fermions at the same temperature and chemical potential given by (S-31).

*Asymptotic momentum distributions.—* The investigation of the large time asymptotics will be performed using the stationary phase approximation. The solution of the Ermakov-Pinney equation in the case of free expansion is $b(t) = (1+\omega_0^2 t^2)^{1/2}$ and in the large time limit we have $\lim_{t\to\infty} b(t) = \omega_0 t$ and $\lim_{t\to\infty} \dot{b}(t) = \omega_0$. The momentum distribution is

$$n_\sigma(p,t) = \sum_{N=1}^\infty \sum_{M=0}^N \sum_{\{j\}} \sum_{\{\lambda\}} p_{N,M}^{\boldsymbol{j},\boldsymbol{\lambda}} \widetilde{G}_{N,M,\sigma}^{\boldsymbol{j},\boldsymbol{\lambda}}(p,t) , \tag{S-43}$$

with $\widetilde{G}_{N,M,\sigma}^{\boldsymbol{j},\boldsymbol{\lambda}}(p,t)$ defined in Eq. (9) of the main text. We are interested in the large $t$ asymptotics of this quantity. If $g(x)$ and $f(x)$ are two reasonably well behaved functions on the real axis with $f'(x_0) = 0$ then the method of stationary phase (Chap. 6 of [S5] or Chap. 2.9 of [S6]) states that

$$\int_\mathbb{R} g(x) e^{itf(x)} dx = \left(\frac{2\pi}{t|f''(x_0)|}\right)^{1/2} g(x_0) e^{itf(x_0)} e^{i\frac{\pi}{4}\mathrm{sign} f''(x_0)} + o\left(\frac{1}{t^{1/2}}\right) , \tag{S-44}$$

when $t\to\infty$. In the large $t$ limit the expression for $\widetilde{G}_{N,M,\sigma}^{\boldsymbol{j},\boldsymbol{\lambda}}(p,t)$ can be written as

$$\widetilde{G}_{N,M,\sigma}^{\boldsymbol{j},\boldsymbol{\lambda}}(p,t) = \frac{b}{2\pi} \int_\mathbb{R} \int_\mathbb{R} G_{N,M,\sigma}^{\boldsymbol{j},\boldsymbol{\lambda}}(\xi_1,\xi_2;0) e^{-itf(\xi_1)} e^{itf(\xi_2)} d\xi_1 d\xi_2 , \tag{S-45}$$

with

$$f(\xi) = \frac{\dot{b}\xi^2}{2l_0^2} - \frac{p\omega_0\xi}{\hbar} . \tag{S-46}$$

The point of stationary phase satisfying $f'(\xi_0) = 0$ is $\xi_0 = p\omega_0 l_0^2/(\dot b\hbar)$. Using (S-44) successively for both integrals in (S-45) we find

$$\widetilde G_{N,M,\sigma}^{\boldsymbol{j},\boldsymbol{\lambda}}(p,t) \underset{t\to\infty}{\sim} \left|\frac{\omega_0 l_0^2}{\dot b}\right| G_{N,M,\sigma}^{\boldsymbol{j},\boldsymbol{\lambda}}\left(\frac{p\omega_0 l_0^2}{\dot b\hbar},\frac{p\omega_0 l_0^2}{\dot b\hbar};0\right). \tag{S-47}$$

From our previous analysis of the initial densities we have $G_{N,M,\downarrow}^{\boldsymbol{j},\boldsymbol{\lambda}}(\xi,\xi) = M G_{N,\mathsf{FF}}^{\boldsymbol{j}}(\xi,\xi)/N$ and $G_{N,M,\uparrow}^{\boldsymbol{j},\boldsymbol{\lambda}}(\xi,\xi) = (N-M)G_{N,\mathsf{FF}}^{\boldsymbol{j}}(\xi,\xi)/N$. The summations in (S-43) are similar with the ones performed in the case of the densities obtaining

$$n_\downarrow^{\mu,B,T}(p,t) \underset{t\to\infty}{\sim} l_0^2\,\frac{e^{B/k_B T}}{2\cosh(B/k_B T)}\,\rho_{\mathsf{FF}}^{\mu',T}\left(\frac{pl_0^2}{\hbar}\right),\quad n_\uparrow^{\mu,B,T}(p,t) \underset{t\to\infty}{\sim} l_0^2\,\frac{e^{-B/k_B T}}{2\cosh(B/k_B T)}\,\rho_{\mathsf{FF}}^{\mu',T}\left(\frac{pl_0^2}{\hbar}\right), \tag{S-48}$$

which shows that the asymptotic momentum distributions have the same shape as the initial densities (property 1 of the main text). Finally, using the identity $n_{\mathsf{FF}}^{\mu,T}(p) = l_0^2\,\rho_{\mathsf{FF}}^{\mu,T}\left(pl_0^2/\hbar\right)$, which is a consequence of the fact that the Hermite functions are eigenfunctions of the Fourier transform (see Appendix E of [S4]) we obtain

$$n_\downarrow^{\mu,B,T}(p,t) \underset{t\to\infty}{\sim} \frac{e^{B/k_B T}}{2\cosh(B/k_B T)}\,n_{\mathsf{FF}}^{\mu',T}(p),\quad n_\uparrow^{\mu,B,T}(p,t) \underset{t\to\infty}{\sim} \frac{e^{-B/k_B T}}{2\cosh(B/k_B T)}\,n_{\mathsf{FF}}^{\mu',T}(p), \tag{S-49}$$

Now, it is easy to see that $n_\downarrow^{\mu,B,T}(p,t) + n_\uparrow^{\mu,B,T}(p,t) \underset{t\to\infty}{\sim} n_{\mathsf{FF}}^{\mu',T}(p)$ which proves the dynamical fermionization at finite temperature (property 2 of the main text).

## IV. PARTITION FUNCTION OF THE GENERAL CASE

In the general case of a system with $\kappa$ components in the presence of a harmonic potential $V(x,t) = \theta(-t)m\omega_0^2 x^2/2$ the Hamiltonian is

$$\mathcal{H} = \int dx\,\frac{\hbar^2}{2m}(\partial_x\Psi^\dagger \partial_x\Psi) + g\,:(\Psi^\dagger\Psi)^2:\,+V(x,t)\,\Psi^\dagger\Psi - \Psi^\dagger\boldsymbol{\mu}\Psi, \tag{S-50}$$

with $\Psi^\dagger = \left(\Psi_1^\dagger(x),\cdots,\Psi_\kappa^\dagger(x)\right)$, $\Psi = (\Psi_1(x),\cdots,\Psi_\kappa(x))^T$ and $\Psi_\sigma(x)$ ($\sigma = \{1,\cdots,\kappa\}$) are fermionic or bosonic fields satisfying the commutation relations $\Psi_\sigma(x)\Psi_{\sigma'}^\dagger(y) - \varepsilon\Psi_{\sigma'}^\dagger(y)\Psi_\sigma(x) = \delta_{\sigma\sigma'}\delta(x-y)$ ($\varepsilon = 1$ in the bosonic case and $\varepsilon = -1$ in the fermionic case). In (S-50) $\boldsymbol{\mu}$ is the chemical potentials matrix with $(\mu_1,\cdots,\mu_\kappa)$ on the diagonal and the rest of the elements 0. In the impenetrable limit, $g\to\infty$, the eigenstates of the system are described by $\kappa$ sets of parameters with $\boldsymbol{j} = \{j_i\}_{i=1}^N$ describing the charge degrees of freedom and $[\boldsymbol{\lambda}] = (\{\lambda_i^{(1)}\}_{i=1}^{N_1},\cdots,\{\lambda_i^{(\kappa-1)}\}_{i=1}^{N_{\kappa-1}})$ with $N \geq N_1 \geq \cdots \geq N_{\kappa-1} \geq 0$ characterizing the spin degrees of freedom. The number of particles in the state $\sigma$ is $m_\sigma = N_{\sigma-1} - N_\sigma$ where we consider $N_0 = N$ and $N_\kappa = 0$ and the eigenstates and will be denoted by $|\Phi^\kappa(\boldsymbol{j},[\boldsymbol{\lambda}])\rangle$. They satisfy $\mathcal{H}|\Phi^\kappa(\boldsymbol{j},[\boldsymbol{\lambda}])\rangle = E_\kappa(\boldsymbol{j},[\boldsymbol{\lambda}])|\Phi^\kappa(\boldsymbol{j},[\boldsymbol{\lambda}])\rangle$ with $|E_\kappa(\boldsymbol{j},[\boldsymbol{\lambda}])\rangle = \sum_{i=1}^N \hbar\omega_0(j_i + 1/2) - \sum_{\sigma=1}^\kappa \mu_\sigma(N_{\sigma-1} - N_\sigma)$.

The wavefunctions are [S7, S8]

$$\chi(\boldsymbol{x}|\boldsymbol{j},[\boldsymbol{\lambda}]) = \sum_{P\in S_N}(\varepsilon)^P P\left[\chi_{\mathsf{FF}}(\boldsymbol{x}|\boldsymbol{j})\theta(x_1 < \cdots < x_N)\otimes\chi_{\mathsf{S}}([\boldsymbol{\lambda}])\right] \tag{S-51}$$

with $\theta(x_1 < \cdots < x_N) = \prod_{j=2}^N \theta(x_j - x_{j-1})$, $\chi_N^{\mathsf{FF}}(\boldsymbol{x}|\boldsymbol{j}) = \det_N[\phi_{j_a}(x_b)]/\sqrt{N!}$ the Slater determinant of free fermions and $\chi_{\mathsf{S}}([\boldsymbol{\lambda}])$ the wavefunction of the relevant spin chain. The correlators can also be written in factorized form like in the Gaudin-Yang model [S9]

$$G_\sigma^{\boldsymbol{j},[\boldsymbol{\lambda}]}(\xi_1,\xi_2) = \langle\Phi^\kappa(\boldsymbol{j},[\boldsymbol{\lambda}])|\Psi_\sigma^\dagger(\xi_1)\Psi_\sigma(\xi_2)|\Phi^\kappa(\boldsymbol{j},[\boldsymbol{\lambda}])\rangle,\quad \sigma = \{1,\cdots,\kappa\},$$

$$= \sum_{d_1}^N\sum_{d_2=1}^N S_\sigma(d_1,d_2)I(d_1,d_2;\xi_1,\xi_2), \tag{S-52}$$

with the charge function independent of the spin

$$I(d_1.d_2;\xi_1,\xi_2) = (-1)^{d_2-d_1}N!\int_{\Gamma_{d_1 d_2}}\prod_{j=2}^N dx_j\,\overline\chi_{\mathsf{FF}}(\xi_1,x_2,\cdots,x_N|\boldsymbol{j})\chi_{\mathsf{FF}}(\xi_2,x_2,\cdots,x_N|\boldsymbol{j}), \tag{S-53}$$

where $\Gamma_{d_1 d_2} = x_2 < \cdots x_{d_1-1} < \xi_1 < x_{d_1} < \cdots < x_{d_2-1} < \xi_2 < x_{d_2} < \cdots < x_N$. The spin functions are defined as

$$S_\sigma(d_1, d_2) = (\varepsilon)^{d_2-d_1} \sum_{\sigma_2,\cdots,\sigma_N} \langle \chi_{\mathsf{S}}([\boldsymbol{\lambda}])|c_{d_1}^\dagger(\sigma)(d_1 \cdots d_2)c_{d_2}(\sigma)|\chi_{\mathsf{S}}([\boldsymbol{\lambda}])\rangle, \tag{S-54}$$

with $c_d(\sigma)$ the annihilation operator of a spin $\sigma$ at the lattice site $d$ and $(d_1 \cdots d_2)$ is a loop permutation operator that cyclically permutes particles and maps $\{d_1, \cdots, d_2\} \to \{d_1+1, \cdots, d_2-1, d_2, d_1\}$. An important relation which is used repeatedly is

$$\sum_{\sigma=1}^\kappa S_\sigma(d,d) = 1, \text{ for all } d \in \{1, \cdots, N\}. \tag{S-55}$$

We want to compute the partition function for a multicomponent system of impenetrable particles in a harmonic potential. We will consider first the case $\kappa = 3$. For a system with three components the eigenstates are indexed by $\boldsymbol{j} = \{j_i\}_{i=1}^N$, $\boldsymbol{\lambda^{(1)}} = \{\lambda_i^{(1)}\}_{i=1}^{N_1}$, $\boldsymbol{\lambda^{(2)}} = \{\lambda_i^{(2)}\}_{i=1}^{N_2}$, and the eigenvalues are (the number of particles in each state is $m_\sigma = N_{\sigma-1} - N_\sigma$ with $N_0 = N$ and $N_3 = 0$)

$$E(\boldsymbol{j}, [\boldsymbol{\lambda}]) = \sum_{i=1}^N \hbar\omega_0(j_i + 1/2) - \mu_1(N - N_1) - \mu_2(N_1 - N_2) - \mu_3 N_2. \tag{S-56}$$

The partition function is

$$\mathcal{Z}(\boldsymbol{\mu}, T) = \sum_{N=0}^\infty \sum_{N_1=0}^N \sum_{N_2=0}^{N_1} \sum_{j_1<\cdots<j_N} \sum_{\lambda_1^{(1)}<\cdots<\lambda_{N_1}^{(1)}} \sum_{\lambda_1^{(2)}<\cdots<\lambda_{N_2}^{(2)}} e^{-\sum_{i=1}^N \frac{\hbar\omega_0(j_i+1/2)}{k_B T}} e^{\frac{\mu_1 N}{k_B T}} e^{\frac{(\mu_2-\mu_1)N_1}{k_B T}} e^{\frac{(\mu_3-\mu_2)N_2}{k_B T}}. \tag{S-57}$$

Taking into account that each $\lambda_i^{(2)}$ ($\lambda_i^{(1)}$) can take $N_1$ ($N$) values and using

$$\sum_{N_2=0}^{N_1} \sum_{\lambda_1^{(2)}<\cdots<\lambda_{N_2}^{(2)}} e^{\frac{(\mu_3-\mu_2)N_2}{k_B T}} = \sum_{N_2=0}^{N_1} C_{N_2}^{N_1} e^{\frac{(\mu_3-\mu_2)N_2}{k_B T}} = \left(1 + e^{\frac{(\mu_3-\mu_2)}{k_B T}}\right)^{N_1}, \tag{S-58}$$

and

$$\sum_{N_1=0}^N \sum_{\lambda_1^{(1)}<\cdots<\lambda_{N_1}^{(1)}} e^{\frac{(\mu_2-\mu_1)N_1}{k_B T}} \left(1 + e^{\frac{(\mu_3-\mu_2)}{k_B T}}\right)^{N_1} = \left(1 + e^{\frac{(\mu_2-\mu_1)}{k_B T}} + e^{\frac{(\mu_3-\mu_1)}{k_B T}}\right)^N, \tag{S-59}$$

we obtain for the partition function for a system with three components

$$\mathcal{Z}(\boldsymbol{\mu}, T) = \sum_{N=0}^\infty \sum_{j_1<\cdots<j_N} \left(e^{\frac{\mu_1}{k_B T}} + e^{\frac{\mu_2}{k_B T}} + e^{\frac{\mu_3}{k_B T}}\right)^N e^{-\sum_{i=1}^N \frac{\hbar\omega_0(j_i+1/2)}{k_B T}}. \tag{S-60}$$

The generalization for arbitrary $\kappa$ is obvious. In the case of pure Zeeman splitting which is described by $\mu_1 = \mu - B(\kappa - 1)$ and $\mu_{i+1} - \mu_i = 2B$ we obtain

$$\mathcal{Z}(\mu, B, T) = \sum_{N=0}^\infty \sum_{j_1<\cdots<j_N} \left(\frac{\sinh(\kappa B/k_B T)}{\sinh(B/k_B T)}\right)^N e^{-\sum_{i=1}^N [\hbar\omega_0(j_i+1/2)-\mu]/k_B T}, \tag{S-61}$$

which shows that the partition function of a gas of trapped impenetrable particles with $\kappa$ components is independent of the statistics of the constituent particles and is the same as the partition of free fermions at the same temperature and renormalized chemical potential

$$\mu'_\kappa = \mu + k_B T \ln[\sinh(\kappa B/k_B T)/\sinh(B/k_B T)]. \tag{S-62}$$

---



\end{document}